\documentclass[aps,prl,twocolumn,showpacs,superscriptaddress,groupedaddress]{revtex4}  
\usepackage{graphicx}  
\usepackage{dcolumn}   
\usepackage{bm}        
\usepackage{amssymb}   
\usepackage{amsmath}
\usepackage{appendix}

\begin{document}

\widetext

\title{The Influence of Decoys on the Noise and Dynamics of Gene Expression}

\author{Anat Burger} \affiliation{Center for Theoretical Biological Physics, University of California, San Diego, La Jolla, CA.}
\author{Aleksandra M.~Walczak} \affiliation{Laboratoire de Physique Th\'{e}orique de l'\'{E}cole Normale Sup\'{e}rieure, Paris, France.  }
\author{Peter G.~Wolynes} \affiliation{Center for Theoretical Biological Physics, Rice University, Houston, TX.}

\date{\today}

\begin{abstract}
{Many transcription factors bind to DNA with a remarkable lack of specificity, so that regulatory binding sites compete with an enormous number of non-regulatory `decoy' sites. For an auto-regulated gene, we show decoy sites decrease noise in the number of unbound proteins to a Poisson limit that results from binding and unbinding. This noise buffering is optimized for a given protein concentration when decoys have a 1/2 probability of being occupied.  Decoys linearly increase the time to approach steady state and exponentially increase the time to switch epigenetically between bistable states.}
\end{abstract}

\maketitle


A transcription factor must bind to a specific site in the genome to regulate the expression of a gene. This process does not occur in isolation. Instead, actual regulatory target sequences must be distinguished from an entire genome of alternative possible binding sites. In prokaryotes, the typical transcription factor binding motif is sufficiently specific that a regulatory target can be distinguished from decoys by its binding free energy alone as a roughly unique location in the genome \cite{g}. Although eukaryotic genomes are much longer, the binding specificity of some eukaryotic transcription factor binding motifs can be so low that up to tens of thousands of strong affinity binding sites can be expected by pure chance \cite{i}. Such decoy sites have been identified in repetitive non-coding regions \cite{t}.  Mutations in these regions have been implicated in several diseases, suggesting that the non-regulatory binding of transcription factors to DNA could serve some currently unknown function, a question that is being explored in synthetically engineered systems \cite{q}.  

A ubiquitous regulatory motif involves a single ``generic'' transcription factor that is responsible for regulating the expression of many genes \cite{v}. As a result, the functional site of one gene may also serve as a decoy site for another gene. Additionally, it is known that active degradation of transcription factors plays an essential role in regulating eukaryotic gene expression. Under certain conditions binding can sterically inhibit--or even prohibit--degradation \cite{c}, so that several eukaryotic transcription factors are protected from degradation while bound to DNA \cite{u}. Previously we have shown \cite{a} that when the bound transcription factors are protected the mean steady state number of $unbound$ transcription factors, $\langle n \rangle$, does not change as decoys are added.  Instead, the $total$ number of transcription factors, $N$, adjusts to satisfy the binding to decoys and thus decoys do not change the deterministic behavior of the system. In this paper we provide an analytical theory of how the noise characteristics and approach to steady state of the system are altered by decoy sites that confer stability through this ``asylum'' mechanism.


\begin{figure}[h]
 \includegraphics[width=\linewidth]{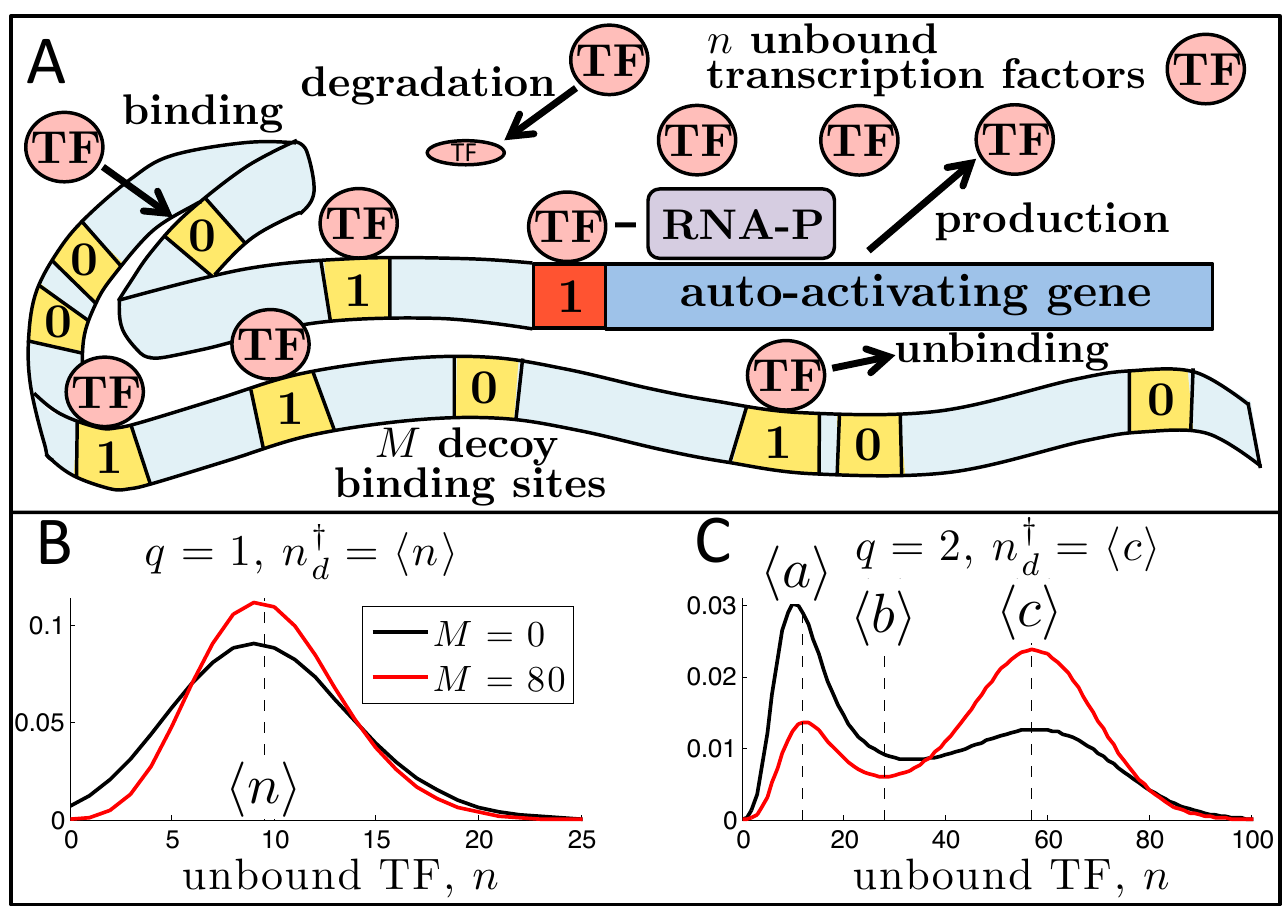}
\caption{{\bf A.} Model of a generic auto-activating gene where transcription factors bind to a regulatory promoter site ({\em red}) as well as $M$ identical, non-regulatory decoy binding sites ({\em yellow}) {\bf B.} Since they protect bound proteins from degradation, decoy binding sites do not alter the steady state mean unbound copy number of a unimodal probability distribution, $\langle n \rangle$, yet they decrease the variance $\sigma_n^2$.  {\bf C.} Similarly, the deterministic fixed points of a bistable system, $\{\langle a \rangle, \langle b \rangle, \langle c \rangle\}$ do not change, but when decoys are added the relative stability of the expression states, LOW ($n<\langle b \rangle$) and HIGH ($n>\langle b \rangle$), is altered.}\label{diagram}
\end{figure}

{\bf The Model.} To elucidate the general effect of decoys on gene expression we model an auto-activated gene surrounded by a collection of $M$ identical decoy binding sites that do not themselves directly regulate transcription but do protect bound proteins from degradation (Fig.~\ref{diagram}). To describe this system we use a master equation (see Appendix A) where each state is described by three indices: the occupancy of the promoter, $i\in\{\text{unbound } (0), \text{bound } (1)\}$, the number of bound decoys, $m$, and the number of unbound proteins, $n$.  Solving this master equation numerically allows us to study properties of the steady state probability distribution over unbound copy numbers, $p_n=\sum_{i,m}p_{i,m,n}(t=\infty)$. The reactions represented in the master equation include protein production $n\xrightarrow{g_i}n+1$, degradation $n\xrightarrow{kn}n-1$, promoter binding, $(i,n)\xrightarrow{H_p(n)(1-i)}(i+1,n-q)$, promoter unbinding, $(i,n)\xrightarrow{f_pi}(i-1,n+q)$, decoy binding, $(m,n)\xrightarrow{H_d(n) (M-m)}(m+1,n-q)$, and decoy unbinding, $(m,n)\xrightarrow{f_dm}(m-1,n+q)$. The binding process encoded in the function $H$ is described for $x\in\{p,d\}$ as $H_x(n)=h_x n$ for binding of monomers ($q=1$) and $H_x(n)=\frac{1}{2}h_x n  (n-1)$ for binding of dimers ($q=2$). We define a site equilibrium constant $n^{\dagger}_x=f_x/h_x$ for $q=1$ and $n^{\dagger}_x=\sqrt{2 f_x/h_x}$ for $q=2$ that corresponds to a binding free energy $E_x$ such that $n^{\dagger}_x=e^{\beta {E_x}}$, where $\beta=(k_BT)^{-1}$. To illustrate the invariant scalings it is convenient to introduce a factor $S$ so that we write the production and promoter binding terms as $g_i=\widehat{g}_iS$ and $n^{\dagger}_p=\widehat{n}^{\dagger}_pS$. This results in $\langle n \rangle=\sum_{n} np_{n}\approx\widehat{\langle n \rangle} S$. The equilibrium probability that a site is occupied is thus a Hill function, $\theta_x(\langle n \rangle)=\langle n\rangle^q/((n^{\dagger}_D)^q+\langle n\rangle^q)$ which can also be written in terms of energy, such that $\theta_x=1/\left(1+\exp\left[\beta q \Delta E\right]\right)$, where $\Delta E=E_x-\mu$ and $\mu=k_BT\ln \langle n \rangle$.  

We focus on the limiting case where binding and unbinding are both much faster than production and degradation; the case of so called ``adiabatic'' genes. We take advantage of the separation in timescales to treat separately the fast fluctuations in unbound copy number--due to binding and unbinding events--from the slow fluctuations in unbound copy number--due to production and degradation events.  The slowly varying component of the unbound copy number, $\overline{n}(N)$, is slaved to a constant total copy number, $N\equiv n+q\cdot i+q \cdot m$, by assuming binding equilibrium. $\overline{n}(N)$ satisfies the equation $N=\overline{n}(N)+q\cdot \theta_p\left[\overline{n}(N)\right]+q\cdot M\cdot \theta_d\left[\overline{n}(N)\right]$.  In this adiabatic limit the full master equation for $p_{i,m,n}$ can be reduced to a one dimensional master equation in terms of the slow variable $N$, $p_N$, by expressing the production and degradation rates as functions of $\overline{n}(N)$ (see Appendix B). This reduced master equation allows us to treat the problem with numerical ease and also find many results analytically.


\begin{figure}
 \includegraphics[width=\linewidth]{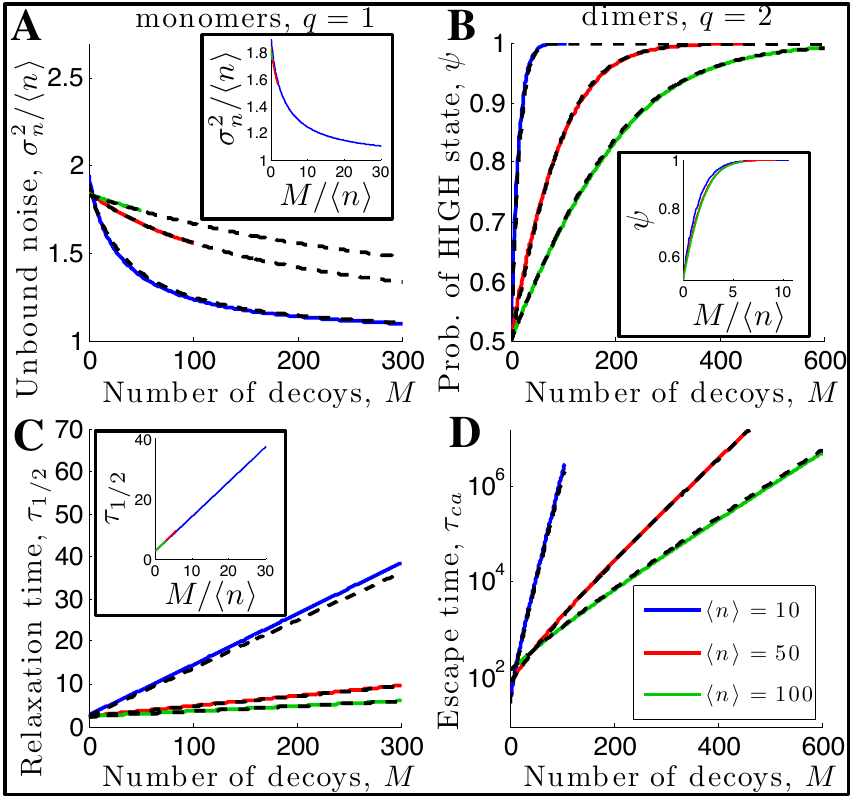}
\caption{Comparison of numerical ({\em solid curves}) and analytical ({\em dashed curves}) results for gene expression properties as decoys are added for systems with varying mean unbound numbers of protein copies, $\langle n\rangle$.  {\bf A.} The Fano factor; {\bf B.} The probability for the bistable system to be in the HIGH protein expression state, $\psi$; {\bf C.} time for the mean total copy number to reach half the steady state value; {\bf D.} epigenetic escape time. Numerical results in {\bf A} are calculated by projecting the solutions of the $3D$ master equation for $p_n=\sum_{i,m} p_{i,m,n}$, whereas the $1D$ master equation for $p_N$ is accurate for the results plotted in {\bf B, C, D}  (see Fig.~\ref{sup4} for details). Numerical calculations for the gene without decoys are used in the analytical calculations.  {\em Parameters:} $g_1=100 S$, $g_0=8S$, $k=1$, $n_P^{\dagger}=53.2 S$ for $q=1$ which gives $\langle n \rangle=50S$.  For $q=2$, $\psi_0=0.5$ is fixed such that $n_P^{\dagger}=10.3$ for $S=.2$, $n_P^{\dagger}=21.0$ for $S=1$, and $n_P^{\dagger}=106.8$ for $S=2$.}\label{trends}
\end{figure}

\begin{figure}
 \includegraphics[width=\linewidth]{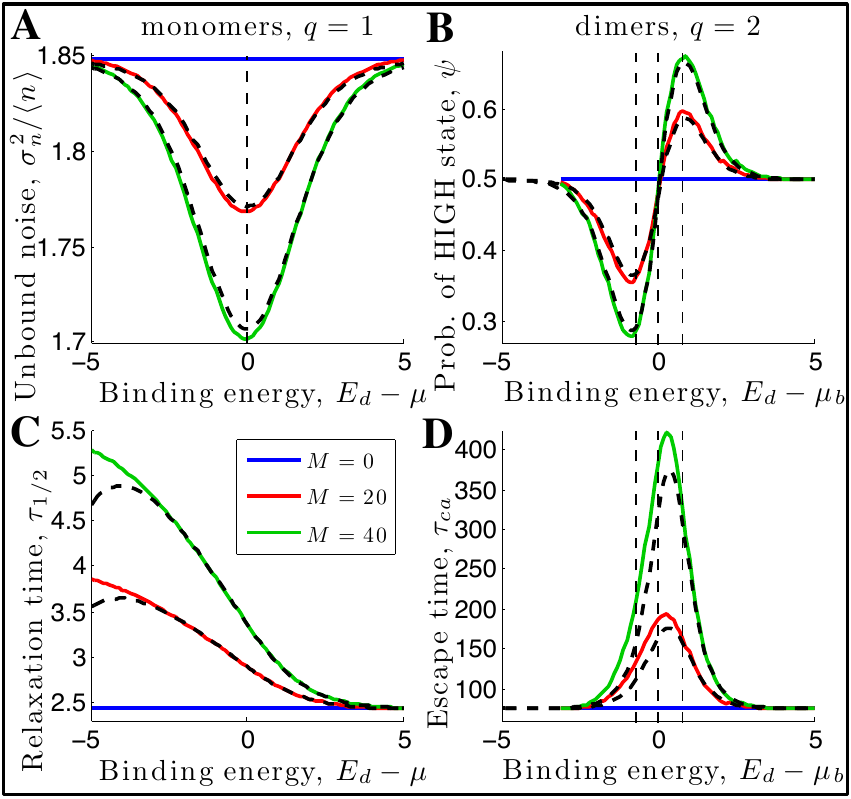}
\caption{Comparison of numerical ({\em solid curves}) and analytical ({\em dashed curves}) for the same properties as in Fig.~\ref{trends} as a function of the decoy binding energy $E_d$, for fixed numbers decoys, $M$. The vertical dashed lines indicate the energies that correspond to the fixed points of the system. Parameters are the same as in Fig.~\ref{trends} for $\langle n \rangle=50$.}
\label{deltaG}
\end{figure}

\begin{figure}[h]
\begin{center}$
\begin{array}{c}
 \includegraphics[width=\linewidth]{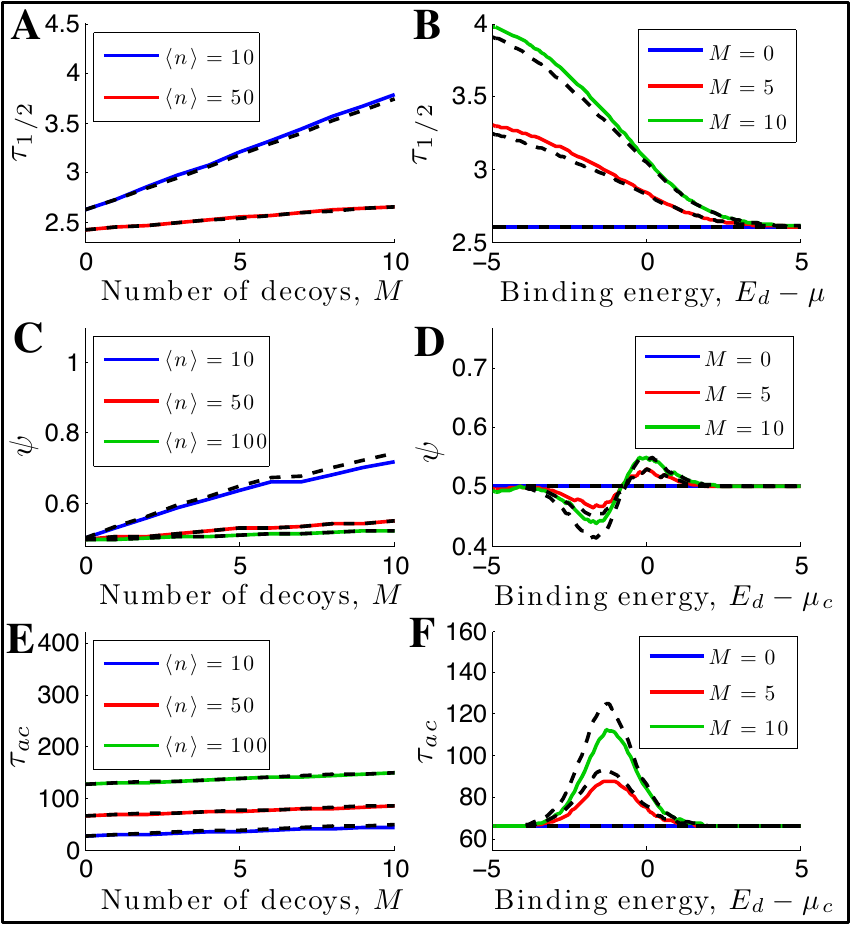}
    \end{array}$
\end{center}
\caption{{\bf Validation of Dimension Reduction.}  Here we compare calculations from the full master equation, Eqn. \ref{nfree}, ({\em colored curves}) with calculations from the one dimensional master equation, Eqn. \ref{oned}, ({\em black dashed lines}).  We see that the dimension reduction breaks down for small system size, $\langle n \rangle$, or strong decoys, $E_d<<\mu$.  {\em Parameters:} $g_1=100\cdot S, g_0=8\cdot S=$ and in panels {\bf A} and {\bf B}, $S=1$, $n^{\dagger}_p=53.2$ $n^{\dagger}_d=10$,  in panels {\bf C}, {\bf D}, {\bf E}, and {\bf F} $ n^{\dagger}_p =10.3$ for $ S = .2, n^{\dagger}_p =21.0$ for $S =1$, and $n^{\dagger}_p$ =106.8 for S =2, with $n^{\dagger}_d=\mu_c$.}
\label{sup4}
\end{figure}

{\bf Numerical Results.} To gain intuition we first numerically solve the master equations for two cases that are known to have qualitatively different dynamical and noise properties without decoys: monomer ($q=1$) and dimer ($q=2$) binding (see caption of Fig~\ref{trends} for details) \footnote{We compare the numerical solutions for the full and reduced model in Fig.~\ref{sup4}.}. Dimer binding allows for bistability and switching between the two attractors, whereas in the adiabatic limit monomer binding yields a unimodal distribution easily characterized by simple measures such as the Fano factor for noise ($\sigma_n^2/\langle n \rangle$) and the mean relaxation time to steady state. In Fig~\ref{trends} we see that adding decoys with a fixed binding energy (we use decoys that are half bound at steady state \footnote{We set  $n^{\dagger}_d=\langle n \rangle$ for $q=1$ and $n^{\dagger}_d=\langle c \rangle$ for $q=2$.}) quantitatively affects the gene expression properties. However when the number of decoys is rescaled by the mean number of unbound proteins, the results for different choices of $S$ collapse onto a common plot (see Fig~\ref{trends} insets) indicating general principles that we explore below. 

We plot the dependence of the noise and dynamical properties of the system on the binding free energy of decoys $E_d$ in Fig~\ref{deltaG}. In prokaryotic genomes, there is typically a free energy penalty of  1 to 2$ k_BT$ per mismatch with respect to the consensus binding motif.  When there are 4 to 5 mismatches  the binding becomes characteristic of background DNA \cite{g}. Since most decoys will have a weaker binding affinity than the promoter, we concentrate on discussing the large $M$,  large $n^{\dagger}_d$ limit \cite{w}.

The steady state unbound Fano factor, $\sigma^2_n/\langle n \rangle$, plotted in Fig.~\ref{trends}A approaches Poisson noise as decoys are added, such that $\sigma^2_n\xrightarrow{M\rightarrow\infty} \langle n \rangle$.  In the limit of large numbers of decoys the slow fluctuations in unbound copy number resulting from production and degradation events are dominated by an effective birth-death process in which a relatively small number of particles bind and unbind to a large reservoir of sites. We see that systems having smaller mean numbers of proteins approach the Poisson limit for smaller values of $M$ (compare blue and green curves in  Fig.~\ref{trends}A) than those with larger mean protein numbers. Figure~\ref{deltaG}A shows that noise buffering is optimized for a particular value of the decoy binding energy, $E_d^*=\mu$.  This corresponds to the case where decoys are half bound at steady state ($n^{\dagger *}_d=\langle n \rangle$). Intuitively, the potential to buffer noise is maximized at $E_d^*=\mu$ since binding and unbinding events are most probable when sites are on average half-occupied.  

Although the mean steady state {\em unbound} protein copy number, $\langle n \rangle$, remains constant, adding decoys increases the mean steady state {\em total} protein number, $\langle N\rangle=\langle n\rangle+M \theta_d(\langle n \rangle)$.  The relaxation time, $\tau_{1/2}$, (the time to reach $\langle N (\tau_{1/2})\rangle=\langle N\rangle/2$, from an initial condition of $\langle N (0)\rangle=0$) increases linearly as decoys are added (Fig~\ref{trends}C) due to the time required to produce the proteins needed to satisfy binding equilibrium. Strongly binding decoys ($E_d<< \mu$) increase $\tau_{1/2}$ the most because more proteins must be created (Fig~\ref{deltaG}C).

In a bistable system where proteins bind as dimers, the addition of decoys does not alter the three deterministic fixed points corresponding to the stable low expression, unstable intermediate expression, and stable high expression levels, $n=\{\langle a \rangle, \langle b \rangle, \langle c \rangle\}$.  However, decoys are able to influence the ability of the system to stochastically transition between the stable global phenotypic states which we call the LOW and HIGH expression states (See Fig~\ref{diagram}C).  The binding affinity of the decoys determines the change in the likelihood of observing the different expression states.  In Fig.~\ref{trends}B we see that decoys with a binding energy $E_d=\mu_c\equiv k_BT\ln\langle c\rangle$ increase the probability to be in the HIGH protein copy expression state by preferentially decreasing fluctuations in the protein buffer in the vicinity of $n=\langle c \rangle$, such that $\psi\xrightarrow{M\rightarrow \infty} 1$ where $\psi=\sum_{n>\langle b\rangle} p_n$.  On the other hand, decoys with a binding energy $E_d=\mu_a\equiv k_BT\ln\langle a\rangle$ will act to stabilize the LOW protein copy number expression state, such that $\psi\rightarrow 0$.  We see that the epigenetic escape times, defined as the mean first passage times between the two steady states, $\tau_{ac}:  n =\langle a \rangle \rightarrow  n=\langle c \rangle$ and $\tau_{ca}: n =\langle c\rangle \rightarrow n=\langle a\rangle$, increase exponentially as decoys are added (Fig.~\ref{trends}D).  The variation of $\psi$ with decoy binding energy (Fig~\ref{deltaG}B) shows that decoys with binding energy $E_d=\mu_b$ stabilize neither state, however, they significantly increase the epigenetic escape rate by effectively stabilizing the transition state (Fig~\ref{deltaG}D).


{\bf Analytical Results.} To understand the numerical observations in Figs.~\ref{trends} and~\ref{deltaG} we approximate the master equation for $p_N$ by a Fokker-Planck equation (see Appendix B). Since the production and degradation rates depend only on the mean unbound protein concentration, $\bar{n}(N)$, the drift and diffusion terms in the Fokker-Planck equation are respectively the sum and difference of the effective production and degradation rates of a gene without decoys evaluated at $\bar{n}(N)$ from the self-consistent relation 

\begin{equation}
N=\bar{n}(N)+qM\theta_d\left[\bar{n}(N)\right].
\label{impl1}
\end{equation}

 To perform the dimensional reduction faithfully it is important to distinguish the unbound concentration given that the promoter is unbound, $\bar{n}$, which determines the probability that the promoter is bound, $\theta_p(\bar{n})$, from the {\em net} unbound concentration, $\bar{n}-q\theta_p(\bar{n})$, which determines the protein degradation rate. Although the reduced model only captures the slow fluctuations in unbound protein concentration we use it to understand the numerical results.  
 
The variance in the unbound protein concentration depends on both fast and slow fluctuations through the law of total variance, $\sigma_n^2=\sigma_{n,slow}^2+\sigma_{n,fast}^2$. The variance from the slow fluctuations, $\sigma_{n,slow}^2$, can be found by taking the small noise approximation of the slow variable Fokker-Planck equation. This result can be rewritten in terms of the mean unbound protein concentration, using a change of variables with the derivative, $\mathcal{J} (\bar{n})\equiv {{\partial N}/{\partial \bar{n}}}$. In the presence of decoys the drift and diffusion functions expressed in terms of $\bar{n}$ are equal to those of a gene without decoys ($M=0$). Thus one finds $\sigma_{n,slow}^2=\sigma_0^2/\mathcal{J}(\langle n \rangle)$, where $\sigma_0^2$ is the variance without decoys (see Appendix B for details). 

To calculate the fast contribution to the variance, $\sigma_{n,fast}^2$, we consider a master equation for the effective birth death process alone that comes from binding and unbinding of the free number of unbound proteins, $n$, to $M$ decoy sites, with a {\em constant} given total number of proteins, $N$ \footnote{We are neglecting promoter binding/unbinding since we are interested in $M>>1$.}. The steady state solution is found by recursion: $p_{n|N}=p_{0|N}\big(n^{\dagger}_d\big)^n\frac{N!}{n!(N-n)!}\frac{(M-N)!}{(M-N+n)!}\equiv\exp\left[\mathcal{F}(n)\right]$.  In the limit of large $N$, $M$, and $n$, within the Stirling approximation we expand $\mathcal{F}(n)$ to second order about its fixed point.  Setting ${\partial_{{n}}\left[ \mathcal{F}({\bar{n})}\right]}=0$ recovers Eqn.~\ref{impl1}.  The variance from this Gaussian expansion of $p_{n|N}$ gives $\sigma_{n|N}^2\approx \bar{n} \left[1-1/\mathcal{J}(\bar{n})\right]$. The fast contribution to the unbound fluctuations is now found by averaging over the probability distributions of the total copy number, $\sigma_{n,fast}^2=\sum_N \sigma_{n|N}^2 p_N\approx \langle n \rangle \left[1-1/\mathcal{J}(\langle n \rangle)\right]$ (see Appendix C for details).  
 
Combining the fast and slow contributions yields
\begin{equation}
\sigma_n^2\approx\Big(\sigma_0^2-\langle n \rangle\Big) \cdot \left[\frac{\left(n^{\dagger}_d+\langle n \rangle\right)^2}{\left(n^{\dagger}_d+\langle n \rangle\right)^2+Mn^{\dagger}_d}\right]+\langle n \rangle\label{one}
\end{equation}

This formula agrees well in the appropriate limits with numerical solutions of the full master equation, as shown in Figs.~\ref{trends}A and \ref{deltaG}A, and also holds for a model that includes translational bursting (see Appendix D). From Eq.~\ref{one} in the large $M$ limit, we obtain the observed Poisson noise, $\sigma_n^2\rightarrow \langle n \rangle $. Noise reduction is proportional to the deviation from Poisson noise in a system without decoys. Decoys will decrease noise for $\sigma_0^2>\langle n \rangle$ \footnote{For a burst size of one, decoys will {\em increase} noise for an auto-repressing gene (where $\sigma_0^2<\langle n \rangle$) and have no effect on a constitutively produced gene (where $\sigma_0^2=\langle n \rangle$).}. Eq.~\ref{one} is minimized for $n^{\dagger*}_d=\langle n \rangle$. Eq.~\ref{one} can be written as a function of $M/\langle n \rangle$ and $\Delta E$ which results in the data collapse shown in the inset of Fig.~\ref{trends}A.

To describe the noise buffering efficacy we quantify the number of decoys needed to reduce the super-Poissonian noise by a half, $M_{1/2}$. We find $M_{1/2}=2\langle n \rangle\left(1+\cosh\Delta E\right)$ is independent of $\sigma_0^2$.  For decoys with optimum buffering capacities ($\Delta E^*=0$), $M_{1/2}=4\langle n \rangle$ and $M_{1/2}$ asymptotically doubles for every binding energy increase of $k_BT\ln 2$ (or doubling of $n^{\dagger}_d$).

The relaxation time to reach $\langle N \rangle/2$ copies of proteins when initially there is no protein present can be obtained directly by integrating the deterministic equation. In the limit of weak decoys, $E_d>\mu$, we find $\tau_{1/2}=\tau_{0,1/2}+M\Delta \tau_{1/2}$, where $\Delta \tau_{1/2}$ is a correction due to decoys, recovering the linear increase of  $\tau_{1/2}$ with decoys seen in Fig.~\ref{trends}C. For very weak decoys, $E_d>>\mu$, (or $n^{\dagger}_d>>\langle n \rangle$), $\mathcal{J}(\bar{n})\approx 1+{M/n^{\dagger}_d}={\rm const}$.   Hence $\Delta \tau_{1/2}\approx \tau_{1/2,0}/n^{\dagger}_d$ (see Appendix E for details). 

Within the Fokker-Planck approximation the epigenetic escape time can be found by expanding the effective potential about the fixed points to second order. In the limit that the barrier height is sufficiently large one finds $\tau_{ac}=\tau_{ac,0}\sqrt{\mathcal{J}(\langle a \rangle)\mathcal{J}(\langle b \rangle)}e^{-M\zeta_{ab}}$, where $\tau_{ac,0}$ is the escape time without decoys and $\zeta_{ab}$ is a correction to the escape path action due to a single decoy. An analogous expression holds for escape from $c$ to $a$. The escape times increase exponentially for large $M$ as decoys are added. 

Since the model has been reduced to one dimension, the bimodal system obeys an effective detailed balance such that $\psi \tau_{ac}=(1-\psi)\tau_{ca}$, where $\psi$ is the probability to be in the HIGH protein copy number expression state.  Using the previous results for the escape times,\\ $\psi=\psi_0\sqrt{\frac{\mathcal{J}(\langle a \rangle)}{{\mathcal{J}(\langle c \rangle)}}}e^{M\zeta_{ac}}/\left(1+\psi_0 \left(\sqrt{\frac{\mathcal{J}(\langle a \rangle)}{{\mathcal{J}(\langle c \rangle)}}}e^{M\zeta_{ac}}-1\right)\right)$.  When $n^{\dagger}\overset{<}{_>}\langle b \rangle$, $\zeta_{ac}\overset{<}{_>}0$ such that $\psi\xrightarrow{M\rightarrow \infty} \overset{0}{_1}$.


In summary, when there is a sufficient separation of timescales between slow protein production-degradation and fast binding-unbinding to the DNA, we have shown that decoys buffer gene expression noise. The fluctuations in binding and unbinding act as an effective birth-death process that imposes the Poisson limit on noise reduction. Noise buffering is optimized for decoys that are half-occupied at the appropriate protein concentration. 

Not all gene regulatory systems function in the fully adiabatic limit explored here \cite{k}. If binding and unbinding to decoys is much slower than the fluctuations in total copy number, decoys are unable to influence the steady state unbound protein expression. If binding and unbinding to the promoter become much slower than the fluctuations in total copy number, there are effectively two gene states with constant production rates.  In this case the decoys have no impact on the steady state unbound protein expression.

{\bf Acknowledgments.} We thank Thierry Mora, Vincent Hakim, and Marc Santolini for helpful discussions, and support from the Center for Theoretical Biological Physics sponsored by the NSF (PHY-0822283), the D.R. Bullard-Welch Chair at Rice University, and the ICAM Junior Travel Grant.


\renewcommand{\theequation}{A\arabic{equation}}    
  \setcounter{equation}{0}  
 
\section{Appendix A: The master equation}\label{apa}

We consider the full master equation for the time evolution of the joint probability distribution of the promoter occupancy ($i \in \{\text{unbound }(0), \text{bound }(1)\}$), the number of occupied decoy sites ($m \in \{0,1,2,..., M\}$) and the number of unbound transcription factors ($n\in\{0,1,2,..., n_{max}\}$): 

\begin{eqnarray}
\partial_t    p_{i,m,n}&=&\Big[g_i   p_{i,m,n-1}-g_i  p_{i,m,n}\Big]\nonumber\\
&&+\Big[k(n+1)  p_{i,m,n+1}-k n   p_{i,m,n}\Big]\nonumber\\
&&+(-1)^{1-i} H_p(n+q i)   p_{0,m,n+q i}\nonumber\\
&&+(-1)^{i} f_p   p_{1,m,n-q (1-i)}\nonumber\\
&&+\Big[H _d(n+q) \Big(M-(m-1)\Big)   p_{i,m-1,n+q}\nonumber\\
&&-H _d(n)\Big(M-m\Big)   p_{i,m,n}\Big]\nonumber\\
&&+f _d\left[\Big(m+1\Big)   p_{i,m+1,n-q}-m   p_{i,m,n}\right].
\label{nfree}
\end{eqnarray}

The reactions (and their rates) that change the state of the system are production ($g_i$) and degradation ($kn$) of transcription factors, binding ($H_p(n)(1-i)$) and unbinding to the promoter ($f_pi$), binding ($H _d(n)(M-m)$) and unbinding to decoys ($f _dm$).  For $x\in\{p,d\}$, transcription factors can either bind as monomers ($q=1$), where $H_x(n)=h_x n$, or as dimers ($q=2$), where $H_x=\frac{1}{2}h_x n (n-1)$. Eq.~\ref{nfree} is solved numerically by matrix diagonalization with $n_{max}>><n>$. 

\section{Appendix B: Slow fluctuations in unbound copy number}\label{apb}

\renewcommand{\theequation}{B\arabic{equation}}    
  \setcounter{equation}{0}  

{\bf Dimensional reduction.} In the limit that binding and unbinding is much faster than production and degradation of transcription factors, the occupancy of the binding sites reaches its steady state on a timescale that is faster than the fluctuations in the {\em total} copy number of transcription factors ($N\equiv n+qi+qm$), which result from production and degradation events.  We collapse the master equation (Eqn. \ref{nfree}) to a single dimension that accounts for effective production, $G(N)$, and degradation, $K(N)$, of proteins:

\begin{eqnarray}
\partial_t   p_{N}&=&\Big[{G}(N-1)  p_{N-1}-{G}(N)  p_{N}\Big]\nonumber\\
&&+\Big[{K}(N+1)  p_{N+1}-{K}(N)  p_{N}\Big].
\label{oned}
\end{eqnarray}

\noindent These rates are defined self-consistently as functions of the slowly varying component of the unbound transcription factors, $\bar{n}$, which depends on the total number of transcription factors, $N$,:

\begin{equation}
N=\bar{n}(N)+q  M  \theta _d[\bar{n}(N)],\label{state}
\end{equation}

\noindent where the probability that a site is bound is given by 

\begin{eqnarray}
{\theta}_x[\bar{n}(N)]=
\begin{cases}
\frac{\bar{n}(N)}{n^{\dagger}_x+\bar{n}(N)}, & \text{if }q=1 \\
\frac{\bar{n}(N)\left[\bar{n}(N)-1\right]}{n^{\dagger}_x (n^{\dagger}_x-1)+\bar{n}(N)\left[\bar{n}(N)-1\right]},  &\text{if }q=2\label{selfcontheta}
\end{cases}
\end{eqnarray}

Equation~\ref{state} does not include a term corresponding to binding to the promoter because the probability that the promoter is bound depends on the mean number of transcription factors {\em given that the promoter is unbound}.  The production rate is a function of the probability that the promoter is bound: 

\begin{equation}
{G}(N)=g_0 (1-\theta_p[\bar{n}(N)])+g_1 \theta_p[\bar{n}(N)]\label{gs}
\end{equation} 

The degradation rate is proportional to the net unbound copy number, which includes the mean number of proteins bound to the promoter:

\begin{equation}
{K}(N)=k  \Big[\bar{n}(N)-q   \theta_p[\bar{n}(N)]\Big].
\label{ks}\end{equation}

{\bf The Fokker-Planck Approximation.} We approximate Eqn. \ref{oned} with a one dimensional Fokker-Planck equation:

\begin{equation}\frac{\partial}{\partial t} p_N = -\frac{\partial}{\partial N}  \left[v(N)-\frac{1}{2} \frac{\partial}{\partial N}  D(N) \right]p_N,\label{fp}
\end{equation}

 \noindent with the drift, $v(N)\equiv G[\bar n(N)]-K[\bar n(N)]$, and diffusion, $D(N)\equiv G[\bar n(N)]+K[\bar n(N)]$.

The steady  state probability distribution of Eq.~\ref{fp} is given by:

\begin{eqnarray}
p(N)=\frac{\mathcal{N}}{{D(N)}}\exp\left[\int_0^{N}dN'\frac{2{v}(N)}{{D}(N)}\right].\label{fpss}
\end{eqnarray}

We define fixed points in total copy number, $N=\{\langle A\rangle, \langle B \rangle, \langle C \rangle\}$, that correspond to the fixed points in unbound copy number, $n=\{\langle a \rangle, \langle b \rangle, \langle c \rangle\}$.  The mean escape time from $N=\langle A \rangle$ to $N=\langle C \rangle$ is \cite{z}:

\begin{equation}
\tau_{AC}={2} \int_{\langle A\rangle}^{\langle C\rangle} {dY} \exp\big[{W}(Y)\big]\int_{0}^Y \frac{dZ}{{D}(Z)} \exp\big[-{W}(Z)\big],\label{tacdef}
\end{equation}

where

\begin{eqnarray}
{W}(N)=-\int_0^{N} dN'  \frac{2 {v}(N')}{{D}(N')}.
\end{eqnarray}

{\bf Small-Noise Approximation.} Within a Gaussian approximation around $N=\langle N \rangle$, Eqn.~\ref{fpss} yields the variance in the total protein copy number:

\begin{equation}
\sigma_{N}^2=\left|\frac{D(N)}{-2{\partial_{N} [v(N)]}}\right|_{N=\langle N \rangle}\label{vardef}.
\end{equation}

One can obtain the variance in the slowly varying component of the unbound protein copies, $\bar{n}$, by performing a change of variables on Equation \ref{vardef} from $N$ to $\bar{n}$.  The drift and diffusion functions evaluated for $\bar{n}$ are equivalent to that of a gene without decoys ($v_0(\bar{n})$ and $D_0(\bar{n})$).  The derivative, $\mathcal{J}(\bar{n})\equiv {\partial N}/{\partial \bar{n}}$, is calculated from Eqn.~\ref{state}:

\begin{eqnarray}
\mathcal{J}(\bar{n})=
\begin{cases}
1+M\frac{n^{\dagger} _d}{(n^{\dagger}_d+\bar{n})^2}, & \text{for }q=1\\
1+M\frac{4(n^{\dagger} _d)^2\bar{n}}{((n^{\dagger}_d)^2+\bar{n}^2)^2}, & \text{for }q=2 
\end{cases}
\end{eqnarray}

\noindent where for dimers we approximate $\bar{n}(\bar{n}-1)\approx \bar{n}^2$.  After the change of variables,

\begin{equation}
\sigma_{n,slow}^2=\left|\frac{D_0(\bar{n})/\mathcal{J}(\bar{n})}{-2{\partial_{\bar{n}} [v_0(\bar{n})]}}\right|_{\bar{n}=\langle n \rangle}=\frac{\sigma_0^2}{\mathcal{J}(\langle n \rangle)},
\end{equation}

\noindent where $\sigma_0^2$ is the variance of the gene without decoys.

Similarly, within a Gaussian approximation about $N=\langle A \rangle$ and $N=\langle B \rangle$, Eqn.~\ref{tacdef} becomes 

\begin{equation}
\tau_{AC}= \frac{2\pi}{D(\langle A\rangle)}\sqrt{\frac{D(\langle A\rangle)D(\langle B\rangle)}{\left|\partial_Nv(\langle A\rangle)\right|\left|\partial_Nv(\langle B\rangle)\right|}}e^{-\int_{\langle A\rangle}^{\langle B\rangle} dN \frac{2 {v}(N)}{D(N)}},\nonumber
\end{equation}

Performing a change of variables from $N$ to $\bar{n}$, the escape time becomes 

\begin{equation}\tau_{ac}=\tau_{ac,0}\sqrt{\mathcal{J}(\langle a \rangle)\mathcal{J}(\langle b \rangle)}e^{-M\zeta_{ab}},
\end{equation}

\noindent where $\tau_{ac,0}$ is the mean escape time without decoys and $\zeta_{ab}$ is the decoy perturbation to the action over the interval $[\langle a \rangle, \langle b \rangle]$:

\begin{equation}
\zeta_{ab}=\int_{\langle a\rangle}^{\langle b\rangle} d\bar{n}'\frac{2 v_0(\bar{n}')}{D_0(\bar{n}')}\left[\frac{4(n^{\dagger} _d)^2\bar{n}'}{((n^{\dagger} _d)^2+\bar{n}'^2)^2}\right].
\end{equation}

Likewise we find $\tau_{ca}=\tau_{ca,0}\sqrt{\mathcal{J}_{2}(\langle c \rangle)\mathcal{J}_{2}(\langle b \rangle)}e^{M\zeta_{bc}}$.

Equivalently, the same formulas for $\sigma_{n,slow}^2$ and $\tau_{ac}$ can be obtained by first performing the change of variables on Eqn.~\ref{fp}, obtaining the effective drift $\widetilde{v}(\bar{n})=v_0(\bar{n}) \mathcal{J}^{-1}(\bar{n})+1/2 D_0(\bar{n}) \mathcal{J}^{-1}(\bar{n})\partial_{\bar{n}}  \mathcal{J}^{-1}(\bar{n})$ and effective diffusion $\widetilde{D}(\bar{n})=D_0(\bar{n}) \mathcal{J}^{-2}(\bar{n})$, followed by the small noise approximation.  

\section{Appendix C: Fast fluctuations in the unbound copy number}\label{apc}

\renewcommand{\theequation}{C\arabic{equation}}    
  \setcounter{equation}{0}  

To study the fast contribution to the variance in the number of unbound protein copies, due to binding and unbinding of monomers, we consider a master equation indexed over the number of unbound transcription factors, $n$, given a {\em constant} total number of transcription factors, $N$.  We neglect binding and unbinding to the promoter because we are interested in the limit of large numbers of decoy sites, $M\rightarrow \infty$. 

\begin{eqnarray}
\frac{dp_{n|N}}{dt}&=&f _d  \Big[(N-n+1)  p_{n-1|N}\nonumber\\
&&-(N-n)  p_{n|N}\Big]\nonumber\\
&+&h _d \Big[(n+1) (M-N+n+1)  p_{n+1|N}\nonumber\\&&-n  (M-N+n)  p_{n|N}\Big]\label{ads}
\end{eqnarray}

The steady state probability distribution is found by recursion:

\begin{eqnarray}
p_{n|N}&=&p_{0|N}  \prod_{\ell=0}^{n-1}\frac{f (N-\ell)}{h  (\ell+1)  (M-N+\ell+1)}\nonumber\\
&=&p_{0|N}  (n^{\dagger})^n \frac{N!}{n! (N-n)!} \frac{(M-N)!}{(M-N+n)!}\nonumber\\
&\equiv& \exp\left[\mathcal{F}(n)\right]\\
&\approx&\exp\left({\mathcal{F}(\bar{n})}\right)  \exp\left[\frac{1}{2}  \left(n-\bar{n}\right)^2  \frac{\partial^2 \mathcal{F}}{\partial n^2}\Big|_{n=\bar{n}}\right].
\end{eqnarray}

In the last step we Gaussian expand $\mathcal{F}$ for large $M$, $N$, and $n$ within a Stirling expansion. Setting $\partial/\partial n \left[\mathcal{F}(\bar{n})\right]=0$ recovers the deterministic result for the mean number of unbound protein copy numbers, $\bar{n}\approx \sum_n n p_{n|N}$  for $\bar{n}>>0$, given in Eqn.~\ref{state}. The variance in the number of unbound protein copy numbers is:

\begin{eqnarray}
\sigma_{n|N}^2&=&\left(\frac{\partial^2 \mathcal{F}}{\partial n^2}\Big|_{n=\bar{n}}\right)^{-1}\nonumber\\&=&\bar{n}  \left[\frac{M  n^{\dagger} _d }{\left(n^{\dagger} _d+\bar{ n}\right)^2+M  n^{\dagger} _d}\right]\nonumber\\
&=&\bar{n}  \left[1-\frac{1}{\mathcal{J}(\bar{n})}\right].
\end{eqnarray}

 The fast contribution to the unbound fluctuations is found by averaging over the probability distributions of the total copy number, $p_N$, which is the steady state solution of Eqn.~\ref{oned} :
 
\begin{eqnarray}
\sigma_{n,fast}^2&=&\sum_N \bar{n}  \left[\frac{M  n^{\dagger} _d}{\left(n^{\dagger} _d+\bar{n}\right)^2+M  n^{\dagger} _d}\right]  p_N\nonumber\\
&\approx&\langle n \rangle    \left[\frac{M  n^{\dagger} _d}{\left(n^{\dagger} _d+\langle n \rangle\right)^2+M  n^{\dagger} _d}\right]\\
&=&\langle n \rangle    \left[ 1-\frac{1}{\mathcal{J}(\langle n \rangle)}\right].\label{analads}
\end{eqnarray}

\noindent We have approximated the average of the function by the function of the average, which is valid for $ \left[\left(n^{\dagger} _d+\langle n \rangle\right)^2+M  n^{\dagger} _d\right]>> 1\nonumber$.

\section{Appendix D: Translational burst noise}\label{ape}

\renewcommand{\theequation}{D\arabic{equation}}    
  \setcounter{equation}{0}  

 Another source of noise in gene expression comes from multiple translation events of a single mRNA copy, so that proteins are effectively produced in bursts rather than one at a time \cite{y}.  Although our model does not include mRNA, we mimic the effects of bursting by specifying that each production event results in an instantaneous burst of $B$ transcription factors with a reduced production rate of transcription factors, $g\rightarrow g/B$, such that the average unbound number of transcription factors $\langle n \rangle$ does not change even though the variance {\em increases}.  For a constitutively produced gene (where $g_0=g_1$) the variance without decoys becomes $\sigma_0^2/\langle n\rangle=(B+1)/2$  \cite{x}. Decoy binding sites have the opposite effect on the variance to bursts -- they {\em decrease} the variance without changing the mean expression $\langle n\rangle$.  The noise buffering formula derived above for $\sigma_n^2=\sigma_{n,slow}^2+\sigma_{n,fast}^2$ can be applied to a constitutively produced bursty gene as follows:
 
\begin{eqnarray}
\sigma_n^2&=&\left(\sigma_0^2-\langle n \rangle\right){\mathcal{J}^{-1}(\langle n \rangle)}+\langle n \rangle\nonumber\\
&=&\langle n\rangle\left[\left(\frac{B-1}{2}\right)\mathcal{J}^{-1}(\langle n \rangle)+1\right].
\end{eqnarray}

There are similar opposing effects between decoys and bursts when one considers the bimodal probability distribution.  Large bursts can eliminate bimodality by decreasing the typical number of production events needed reach the transition state from a fixed point \cite{e}, such that the probability of the HIGH state decreases.  Adding decoys that stabilize the HIGH state ($n^{\dagger}_d >\langle b \rangle$) can restore bimodality in a bursty bimodal system.  Similarly, bursts exponentially decrease the time to escape between states \cite{f}, whereas decoys exponentially increase the time to escape between states.

\section{Appendix E: Approach to steady state}\label{apd}

\renewcommand{\theequation}{E\arabic{equation}}    
  \setcounter{equation}{0}  

The time to reach half of the mean steady state expression,  $\tau_{1/2}$,  starting from a mean of zero protein copies is found from the deterministic equation for the mean total copy number, $d_t\langle N (t)\rangle=v(N)=v_0\left[\bar{n}(N)\right]$, to be:

\begin{equation}
\tau_{1/2}=\int_0^{\langle N\rangle/2}dN\frac{1}{v_0\left[\bar{n}(N)\right]}.\label{taudef}
\end{equation}

Performing a change of variables from $N$ to $\bar{n}$ yields:

\begin{equation}
\tau_{1/2}=\int_0^{\bar{n}(\langle N\rangle/2)}d\bar{n}\frac{\mathcal{J}(\bar{n})}{v_0(\bar{n})}
\end{equation}

\noindent where the upper boundary is the mean unbound copy number $\bar{n}$ such that Eqn. \ref{state} is evaluated for $N=\langle N \rangle/2$ for binding of monomers.

{\bf Limit of weak decoys}. For weak decoys ($n^{\dagger}_d>>\langle n \rangle$), approximating $\theta_d(\bar{n})\approx {\bar{n}}/{n^{\dagger}_d}$ in Eqn \ref{state} results in $N\approx\bar{n} (1+M/n^{\dagger}_d)$ and $\mathcal{J}(\bar{n})=\partial{N}/\partial{\bar{n}}=1+M/n^{\dagger}_d=const.$  In this limit the upper boundary becomes $\bar{n}(\langle N\rangle/2)\approx \langle n \rangle/2$ and $\tau_{1/2} = \tau_{1/2,0}+M\Delta\tau_{1/2}$ where $\Delta \tau_{1/2}\approx \tau_{1/2,0}/n^{\dagger}$.

{\bf Limit of strong decoys.} For strong decoys ($n^{\dagger}_d<<\langle n \rangle$),  Eqn. \ref{state} becomes $N\approx\bar{n}+M(1-n^{\dagger}/\bar{n})$, and $\mathcal{J}(\bar{n})\approx 1+ Mn^{\dagger}_d/\bar{n}^2$.  Therefore, unlike weak decoys that influence $\tau_{1/2}$ independently of $\bar{n}$, strong decoys have the most significant effect of increasing the time to reach the steady state (compared to the gene with no decoys) when $\bar{n}$ is small.  

In the limit of extremely strong decoys, each transcription factor that is produced binds to a decoy site and remains bound.  As a result, until all decoys are saturated, the unbound copy number will be zero. There will be no transcription factors available to bind to the promoter and the production will be fixed at the basal production level, $g_0$.  After saturation, however, strong decoys no longer influence the dynamics of the system.  Therefore the time to approach steady state can be broken up into a basal production stage and an isolated gene stage.
 
 For $M > \langle n \rangle$, the time to reach half of the steady state number of proteins happens before the decoys are saturated - in the regime when transcription factors are produced with a rate $g_0$ per unit time,

\begin{equation}
\tau_{1/2}=\frac{\langle N \rangle}{2\cdot g_0}=\frac{M+\langle n \rangle}{2\cdot g_0},\text{ for } M >\langle n \rangle>>n^{\dagger}_d.\label{ts1}
\end{equation}

\end{document}